\documentclass[runningheads]{llncs}
\usepackage[T1]{fontenc}
\usepackage{graphicx}
\usepackage{hyperref}
\usepackage{color}

\renewcommand{\arraystretch}{1}
\usepackage[margin=1.2in]{geometry}

\usepackage{caption}
\usepackage{bbm}
\usepackage{algorithm}
\usepackage{algpseudocode}
\usepackage{amsmath}
\usepackage{amsfonts}
\usepackage{comment}
\usepackage{tcolorbox}
\tcbuselibrary{skins,breakable}
\usepackage[utf8]{inputenc}
\usepackage{xcolor}
\usepackage{listings}
\usepackage{graphicx}
\usepackage{subfig}
\usepackage{booktabs}
\usepackage{array}
\usepackage{xcolor}
\usepackage[table]{xcolor}
\usepackage{float}
\usepackage[multiple]{footmisc}
\usepackage{makecell}

\usepackage{fvextra}

\begin{document}
\title{GPU Acceleration of Learning With Errors KEMs Using OpenACC for Post-Quantum Cryptography\thanks{Pre-print version of the manuscript submitted to NUMTA2026}}
\titlerunning{GPU Acceleration of Plain LWE KEMs}
%
\author{
    Tiziana Liberati\inst{1,3}\and
    Nitin Shukla\inst{2}\and
    Matteo Barbieri\inst{1,4,5}\and
    Gabriella Bettonte\inst{1}\and
    Elisabetta Boella\inst{1}\and
    Simone Rizzo\inst{1}\and
    Daniele Gregori\inst{1}\and
    Marco Pedicini\inst{3}
}
\authorrunning{T. Liberati et al.}
%
\institute{
E4 Computer Engineering SpA, via Martiri della Libertà 66, 42019 Scandiano (Italy) 
\email{\{tiziana.liberati, matteo.barbieri, gabriella.bettonte, elisabetta.boella, simone.rizzo, daniele.gregori\}@e4company.com}
\and
SuperComputing Applications and Innovation Department, Cineca, via Magnanelli 6/3, 40033 Bologna (Italy)
\email{n.shukla@cineca.it}
\and
Dipartimento di Matematica e Fisica, Roma Tre University, Largo San Leonardo Murialdo 1, 00146 Rome (Italy)
\email{marco.pedicini@uniroma3.it}
\and
Dipartimento di Fisica e Astronomia “G. Galilei”, Universit\`a di Padova, via F. Marzolo 8, 35131 Padova (Italy)
\and INFN, Sezione di Padova, Italy
}
\maketitle              
\begin{abstract}
Shor's algorithm proved that asymmetric cryptographic protocols based on the integer factorization and discrete logarithm problems are no longer safe in a world with large-scale quantum computers. As a result, Post-Quantum Cryptography (PQC) has been developed over the last few years, seeking cryptographic primitives resistant to quantum attacks. One of the main hard problems underlying PQC schemes is the Learning with Errors (LWE) problem, which is significantly more computationally intensive than its classical predecessors. 
In this work, we present a Key Encapsulation Mechanism (KEM) based on plain LWE and develop a GPU-oriented implementation using OpenACC. We evaluate the performance of our accelerated application in terms of both time-to-solution and energy-to-solution, considering bare-metal and containerized executions across multiple NVIDIA GPU models and generations. 
Our implementation achieves significant acceleration across all tested GPU platforms.
In particular, on the NVIDIA Grace Hopper Superchip, it attains up to a $208\times$ speedup over a multithreaded CPU baseline and enables the execution of problem sizes that are impractical on CPU architectures due to memory and synchronization constraints.
Energy consumption analysis also shows $\approx 2\times$ better efficiency when using the Superchip compared to systems equipped with x86-based CPUs and NVIDIA H100 GPUs.
These results highlight the effectiveness of GPU acceleration for computationally demanding LWE-based cryptographic workloads.

\end{abstract}

\section{Introduction}
Shor's algorithm~\cite{Shor} solves the integer factorization and the discrete logarithm problems in polynomial time with a quantum computer, threatening widely used public-key cryptosystems, such as Rivest–Shamir–Adleman (RSA) and Elliptic-Curve Diffie-Hellman. In response, the National Institute of Standards and Technology (NIST) has initiated standardization of Post-Quantum Cryptography (PQC) protocols. In this realm,  lattice-based schemes are widely studied due to their security and flexibility, particularly those based on Learning With Errors (LWE) for key encapsulation mechanisms (KEMs)~\cite{FIPS203}. Albeit being secure from known quantum attacks, they introduce significant computational overhead from large-scale matrix operations and randomness generation~\cite{KUMAR2022100242}.

To address these limitations, PQC schemes have been accelerated on GPUs leveraging their parallel processing capabilities, particularly in lattice-based constructions. Implementations of LWE-based schemes such as FrodoKEM, NewHope, and Kyber~\cite{PQC-KEM-GPU}, as well as other lattice-based cryptographic primitives~\cite{Dilithium,PQC-KEM-GPU}, have demonstrated substantial speedups through optimized memory access patterns, efficient data transfer, and data-parallel execution strategies. However, most existing approaches rely on CUDA-specific optimizations.
In this work, we propose a KEM based on plain LWE and investigate a directive-based heterogeneous programming approach using OpenACC.
This method enables existing CPU code bases to leverage GPUs with minimal modifications, thereby improving developer productivity while preserving code readability and long-term maintainability.
Although OpenACC currently provides more limited portability than alternative heterogeneous programming models, it offers a favorable trade-off between programmability and performance~\cite{Stack2021}. Moreover, the maturity of the NVIDIA compiler software stack now enables high performance, in some cases approaching or even exceeding CUDA~\cite{Vignolo2025}. Furthermore, thanks to the straightforward interoperability between OpenACC and CUDA~\cite{boella}, we effortlessly integrate a CUDA-based modular Advanced Encryption Standard (AES)-based random number generator (RNGonGPU\footnote{RNGonGPU library Git repository: \url{https://github.com/Alisah-Ozcan/RNGonGPU/}}) to produce noise directly on the GPU, reducing CPU-side overhead during the main KEM operations and thus mitigating one of the primary bottlenecks in LWE-based cryptosystems. 
In addition, we introduce an intra-operation batching strategy that exploits fine-grained parallelism within KEM operations by supporting bit-level parallel execution. Unlike inter-instance batching used in prior work, this method improves resource utilization within a single KEM instance while mitigating kernel-level inefficiencies.

Overall, our implementation achieves speedups of up to $208\times$ compared to the CPU OpenMP version, mainly due to the parallelization of the matrix--vector multiplications. Importantly, it also enables the efficient handling of problem sizes that are impractical to process on CPUs.

Besides comparing against the CPU implementation, this work also evaluates the performance of the proposed approach across multiple GPU generations and models, analyzing how architectural characteristics influence execution time and energy consumption. 

Thus, the main contributions of this paper are:
\begin{itemize}
    \item An open-source implementation of a plain LWE-based KEM for CPU and an OpenACC accelerated version, enabling reproducible experiments and benchmarks.
    \item A set of GPU-oriented optimizations, applicable to a broad class of LWE-based cryptographic schemes, along with an evaluation and discussion of the main optimization strategies used to address the primary performance bottlenecks.
    \item A comprehensive performance, energy-efficiency, and scalability study across different parameter sets, problem size, and NVIDIA GPU architectures.
\end{itemize}

This paper is organized as follows. Section~\ref{sec:implementation} presents implementation details of the post-quantum cryptosystem, with emphasis on its parallelization using OpenMP on multicore architectures and on the choice of valid parameter sets. Section~\ref{sec:gpu} describes the strategy adopted to extend the implementation with GPU support. 
Section~\ref{sec:exp} reports performance, energy consumption, and scalability results for the proposed implementation across multiple NVIDIA GPU architectures.
Finally, Section~\ref{sec:confun} summarizes the main findings and discusses their implications for high-performance GPU software development.
All performance evaluations and benchmark reproductions in this work are based on the implementation hosted in the \emph{LWE-KEM} Git repository\footnote{LWE-KEM Git repository: \url{ https://github.com/TizianaLiberati/LWE-KEM/}.}.

\section{Implementation and Parallelization of the Post-Quantum Cryptosystem}
\label{sec:implementation}

This section introduces the mathematical and cryptographic foundations underlying the proposed plain LWE KEM.

We first briefly review the security assumptions of lattice-based cryptography, which motivate the large adoption of lattice problems in modern post-quantum cryptographic schemes.
We then introduce the notion of Key Encapsulation Mechanisms (KEMs) together with the underlying public-key encryption construction on which KEMs rely. Subsequently, we present a KEM based on the LWE problem and the main algorithms composing the scheme.

We also discuss the parameter choices adopted in this work and the validation methodology used to ensure correct decapsulation behavior across both CPU and GPU implementations.

Finally, we consider the OpenMP multicore parallelization and highlight the bottlenecks of the scheme; motivating the adoption of GPU acceleration strategies that will be analyzed later on.

\subsection{Mathematical Framework}
The proposed KEM is based on the Learning With Errors (LWE) problem. Intuitively, LWE can be viewed as solving noisy linear system modulo $q$: the small error term $e$ hides the linear relation between the public matrix $A$ and the secret $s$, making the recovery of the secret computationally hard.
Formally, in the LWE setting, given a random matrix $A \in \mathbbm{Z}^{n \times n}_q$, a secret vector $s \in \mathbbm{Z}_q^n$, and a small error vector $e \in \mathbbm{Z}_q^n$, where $n$ is the lattice dimension, one observes samples of the form
\begin{equation}\label{eq:LWE}
    b = As + e \pmod q.
\end{equation}
The security of the scheme relies on the difficulty of either recovering $s$ from such samples (search-LWE) or distinguishing valid samples from random values (decision-LWE)~\cite{Regev}.
LWE is a central hardness assumption in lattice-based cryptography due to its worst-case to average-case reductions~\cite{Regev}, which relate its difficulty to lattice problems such as GapSVP, BDD and SIVP. 

These hardness assumptions are the foundation on which many post-quantum protocols are built.
During the NIST standardization process for PQC, two main primitives were considered: digital signatures and key encapsulation mechanisms.
In this work, we focus on KEM constructions, which are a type of protocols that enable secure establishment of shared symmetric keys over untrusted channels. 
A KEM can be constructed from an underlying public-key encryption (PKE) scheme as shown in Figures~\ref{fig:algos1} and~\ref{fig:algos2}. It is composed of three main algorithms: \textit{KeyGen}, which generates the secret and the public key ($sk$ and $pk$) (Algorithm~\ref{alg:keygen}); \textit{Encrypt}, which encrypts the message $m$ (a pseudorandom bit string) into a ciphertext $c$ (Algorithm~\ref{alg:encrypt}); \textit{Decrypt} which recovers the encrypted message $m$ from the ciphertext (Algorithm~\ref{alg:decrypt}). The KEM extends this PKE protocol through \textit{Encaps}~(Algorithm~\ref{alg:encaps}) and \textit{Decaps}~(Algorithm~\ref{alg:decaps}). \textit{Encaps} generates both the ciphertext and the shared secret key, while \textit{Decaps} decrypts the ciphertext, recomputes it deterministically, and verifies its correctness through re-encryption. This mechanism is implemented using the Fujisaki-Okamoto (FO) transform which upgrades the scheme to IND-CCA2 security~\cite{FO}. If the recomputed ciphertext matches the received one, the shared secret is derived from the valid pair (Figure~\ref{fig:kem}); otherwise, implicit rejection outputs a fallback pseudorandom value from the secret key and ciphertext, ensuring constant-time decapsulation without leaking information about ciphertext validity. In our implementation, this appears as: (i) explicit re-encryption verification, (ii) hash-based fallback secret derivation, and (iii) a constant-time decapsulation path independent of ciphertext validity.
\begin{figure}[t]
\centering

\begin{minipage}[t]{0.32\textwidth}
\begin{algorithm}[H]
\caption{\texttt{KeyGen}$()$}
\label{alg:keygen}
\begin{algorithmic}[1]
\State $A \gets \texttt{sample}(\rho) \in \mathbbm{Z}_q^{n \times n}$
\State $s \gets \chi_\eta^n$
\State $e \gets \chi_\eta^n$
\State $t \gets A s + e \pmod q$
\State \Return $pk=(A,t),\ sk=s$
\end{algorithmic}
\end{algorithm}
\end{minipage}
\hfill
\begin{minipage}[t]{0.32\textwidth}
\begin{algorithm}[H]
\caption{\texttt{Encrypt}$(pk,m)$}
\label{alg:encrypt}
\begin{algorithmic}[1]
\State $e_2 \gets \mathcal{U}_k$
\State $r \gets \chi_\eta^n$
\State $e1 \gets \chi_\eta^n$
\State $u \gets A^T r + e_{1} \pmod q$
\State $v \gets t^T r + e_{2} + m \pmod q$
\State \Return $c=(u, v)$
\end{algorithmic}
\end{algorithm}
\end{minipage}
\hfill
\begin{minipage}[t]{0.32\textwidth}
\begin{algorithm}[H]
\caption{\texttt{Decrypt}$(sk,c)$}
\label{alg:decrypt}
\begin{algorithmic}[1]
\State $\mu \gets v - s^T u \pmod q$
\If{$\mu_j \le q/4$ or $\mu_j \ge 3q/4$}
    \State $m_j \gets 0$
\Else
    \State $m_j \gets q/2$
\EndIf
\State \Return $m$
\end{algorithmic}
\end{algorithm}
\end{minipage}
\caption{Pseudo-code of the underlying LWE-based public key encryption scheme, showing key generation, encryption, and decryption algorithm.}
\label{fig:algos1}
\end{figure}

\begin{figure}[t]
\centering

\begin{minipage}[t]{0.48\textwidth}
\begin{algorithm}[H]
\caption{\texttt{Encaps}$(pk)$}
\label{alg:encaps}
\begin{algorithmic}[1]
\State $m \gets \{0,1\}^{256}$
\State $c \gets \texttt{Encrypt}(pk,m)$
\State $K' \gets H(H(pk) \,\|\, m)$
\State $K \gets H(K' \,\|\, H(c))$
\State \Return $(c,K)$
\end{algorithmic}
\end{algorithm}
\end{minipage}
\hfill
\begin{minipage}[t]{0.48\textwidth}
\begin{algorithm}[H]
\caption{\texttt{Decaps}$(sk,c)$}
\label{alg:decaps}
\begin{algorithmic}[1]
\State $m' \gets \textsc{Decrypt}(sk,c)$
\State $c' \gets \textsc{Encrypt}(pk,m')$
\State $K' \gets H(H(pk) \,\|\, m')$
\If{$c'=c$}
    \State $K \gets H(K' \,\|\, H(c))$
\Else
    \State $K \gets H(z \,\|\, H(c))$
\EndIf
\State \Return $K$
\end{algorithmic}
\end{algorithm}
\end{minipage}
\caption{Pseudo-code of the LWE-based key encapsulation mechanism, showing encapsulation and decapsulation using the Fujisaki--Okamoto transform.}
\label{fig:algos2}
\end{figure}

\begin{figure}[h]
    \centering
    \includegraphics[width=0.6\linewidth]{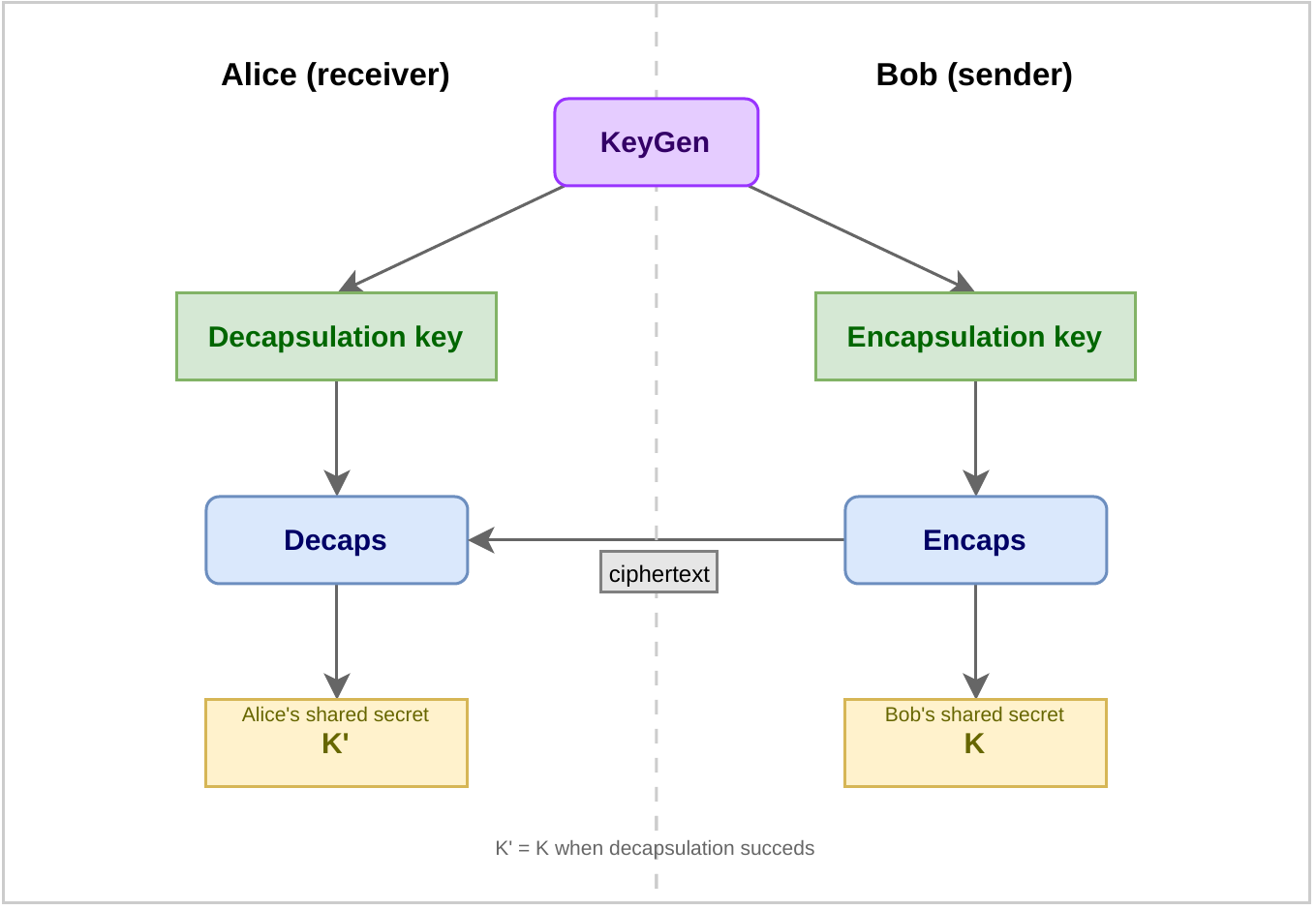}
    \caption{Key establishment using a KEM}
    \label{fig:kem}
\end{figure}

\subsection{Correctness and Parameter Validation}

The correctness of the proposed KEM is determined by the choice of the LWE parameters, namely the lattice dimension $n$, the modulus $q$, and the adopted noise distributions. Intuitively, decryption succeeds only when the accumulated noise remains sufficiently small compared to the modulus; otherwise, the recovered value may exceed the decoding threshold and lead to an incorrect message reconstruction.

In this work, the secret key distribution is fixed to a binomial distribution with parameter $\eta=3$, a standard and well-studied choice in lattice-based cryptography~\cite{FIPS203}. The remaining parameters $(n, q, \sigma, k)$ are varied to balance decryption correctness and computational hardness. Table~\ref{tab:parameters} displays the parameter sets considered in our evaluation, including lattice dimension, modulus, noise parameters, and the corresponding approximate hardness.

The selected noise distributions follow common practices in LWE-based schemes, providing well-understood hardness properties while preserving reliable decryption behavior, and their approximate hardness was assessed using the fast LWE parameter selection tool by Biasoli et al.~\cite{Marcolla}, under simplified LWE model with binary secrets and Gaussian noise. The reported values in Table~\ref{tab:parameters} should therefore be interpreted as indicative hardness levels, rather than exact KEM bit-security claims.

\begin{table}[htp!]
\caption{Selected LWE parameters and estimated hardness (i.e., security level)}
\vspace{1mm}
\label{tab:parameters}
\centering
\small

\setlength{\tabcolsep}{3pt}
\renewcommand{\arraystretch}{1.0}

\begin{tabular}{c c c c c c}
\toprule
\textbf{$n$} & \textbf{$q$} 
& \makecell{\textbf{Gaussian $\sigma$}\\ \textbf{($e,r,e_1$)}} 
& \makecell{\textbf{Centered uniform $k$}\\ \textbf{($e_2$)}} 
& \makecell{\textbf{Binomial $\eta$}\\ \textbf{($s$)}} 
& \makecell{\textbf{Estimated}\\ \textbf{hardness}} \\
\midrule

512    & 3329  & 2.3 & 3 & 3 & 141 \\
1024   & 12289 & 2.3 & 3 & 3 & 257 \\
2048   & 12289 & 2.3 & 3 & 3 & 559 \\
4096   & 12289 & 1.7 & 3 & 3 & 1230 \\
8192   & 65537 & 1.2 & 2 & 3 & 2323 \\
16384  & 65537 & 1.2 & 2 & 3 & 5070 \\
32768  & 65537 & 1.0 & 2 & 3 & 10990 \\
65536  & 65537 & 1.0 & 1 & 3 & 23699 \\
131072 & 65537 & 1.0 & 1 & 3 & 50855 \\

\bottomrule
\end{tabular}
\label{tab:parameters}
\end{table}

Overall, the chosen parameter sets are intended to ensure reliable decapsulation in our experiments while spanning a range of increasing, approximate hardness levels.

\subsection{Bottlenecks and Parallelization Approach}
\label{bottlenecks}

Our baseline implementation is written in C++ and parallelized using OpenMP to exploit multicore architectures, addressing the high computational cost of lattice-based post-quantum cryptographic schemes. For LWE-based KEMs, the dominant contribution to the overall computational cost arises from the matrix--vector multiplications in Equation~\eqref{eq:LWE}. This operation is executed in all three KEM phases (\textit{KeyGen} Algorithm~\ref{alg:keygen}, \textit{Encaps} Algorithm~\ref{alg:encaps}, and \textit{Decaps} Algorithm~\ref{alg:decaps}) and dominates runtime due to the quadratic memory traffic associated with the public matrix $A$ in Equation~\eqref{eq:LWE}. Additional costs arise from dot-product reductions, matrix transposition, and the generation of random noise. In our plain LWE-KEM additional overhead is introduced by processing 256 independent message bits separately in both encapsulation and decapsulation.

As the problem size increases, these costs lead to  high latency and limited scalability, especially in server-side settings with many concurrent key exchanges, where the CPU becomes a throughput bottleneck.
At the same time, these operations (matrix generation, matrix--vector multiplication, dot products, per-bit ciphertext processing, and noise generation) are inherently data-parallel making them well suited for parallel execution. Therefore, our optimization strategy begins with shared-memory parallelism using OpenMP and is later extended to GPU offloading via OpenACC, to better exploit large-scale thread parallelism and memory bandwidth.

\section{GPU Porting and Optimization of LWE-KEM}
\label{sec:gpu}

This section presents our OpenACC-based GPU implementation of the proposed plain LWE-KEM. We begin from a naive OpenACC baseline, in which the relevant computational loops are offloaded to the GPU while random numbers are generated on the host and subsequently transferred to the device. Building upon this baseline, we progressively introduce a series of refinements, including intra-operation batching, device-resident random number generation, explicit data management, and kernel parallelization strategies. Before presenting these techniques, we motivate the main design choices and identify the primary performance bottleneck. The section concludes with a detailed performance analysis. All performance results reported in this section were obtained on a Booster node of CINECA's \textit{Leonardo} supercomputer, whose hardware and software specifications are provided in Appendix~\ref{sec:appendix}.


\subsection{Design Rationale}
The optimization strategy is driven by the bandwidth asymmetry between host and device memory systems. PCIe Gen4 provides approximately 32\,GB/s bandwidth, whereas NVIDIA A100 global memory bandwidth reaches approximately 1.5\,TB/s. This yields an effective ratio of approximately 50:1. Three architectural principles follow from this observation:
\begin{enumerate}
    \item \textbf{Device-resident random number generation}: all randomness is
          generated directly on the GPU, eliminating bulk host-to-device
          transfer.
    \item \textbf{Explicit data management}: OpenACC \texttt{enter data} and
          \texttt{exit data} directives are used for deterministic control over
          host-device transfers.
    \item \textbf{Batched kernel execution}: all 256 component encryptions are
          fused into a single kernel launch to amortize invocation overhead.
\end{enumerate}

In addition, the public matrix $A \in \mathbbm{Z}_q^{n \times n}$ and the associated working buffers are managed to minimize unnecessary host-device communication and temporary allocation overhead during large-scale executions.

\subsection{Intra-Operation Batching}
A significant inefficiency in the naive OpenACC baseline originates from sequential processing of 256 independent encryption instances. Each instance triggers separate memory transfers and kernel launches. 
The baseline execution model is characterized as follows:

\[
\texttt{CPU noise generation}
\rightarrow
\texttt{Host-to-device transfer}
\rightarrow
\texttt{GPU encryption}
\]
and repeated independently for each of the 256 message bits.

This design introduces excessive synchronization and launch overhead. To mitigate this limitation, an intra-operation batching strategy is introduced. All noise vectors are precomputed and transferred in a single host-to-device operation.

\[
\begin{aligned}
    \texttt{CPU seed generation}
\rightarrow
\texttt{Host-to-device transfer}
\rightarrow
\texttt{Noise generation on GPU} \\
\rightarrow
\texttt{Single batched GPU kernel for 256 encryptions}
\end{aligned}
\]

For $n = 4096$, where $n$ is the lattice rank, the batched noise representation includes $256 \times n$ elements. This corresponds to approximately 8\,MB per encapsulation.

The GPU kernel is executed once per batch, reducing launch overhead from 256 invocations to a single invocation. The resulting execution exposes $256 \times n$ independent work items that can be executed concurrently by the OpenACC implementation.. 

\subsection{Device-Resident Random Number Generation}
For $n = 4096$, each encapsulation requires approximately $2.1 \times 10^6$ random samples. 
Transferring these as 32-bit integers over PCIe Gen4
($\sim$16\,GB/s effective) introduces approximately 0.5\,ms of latency per call, which is comparable to the GPU compute time itself.
Moreover, CPU-sequential generation places noise generation on the critical path, leaving the device idle while randomness is produced.

To eliminate both the transfer and serialization bottlenecks,we employ the RNGonGPU library for fully on-device pseudorandom sampling. Since RNGonGPU implements an AES-256 Deterministic Random Bit Generator (DRBG) compliant with NIST SP\,800-90A, it provides cryptographic-grade randomness together with forward secrecy and backtracking resistance. Consequently, only a 64-bit seed is transferred from host to device at initialization, while all subsequent pseudorandom expansion is performed on-device. This reduce the per-encapsulation host-to-device transfer volume by approximately six orders of magnitude (from $\sim$8\,MB to 8\,bytes).
As a result, this produces a throughput on the NVIDIA A100 that exceeds $10^9$\,samples/s, compared to $\sim 10^7$\,samples/s for sequential OpenSSL, results in a two-order-of-magnitude advantage. Furthermore, RNGonGPU generates identical pseudorandom sequences across GPU architectures (Volta, Ampere, Hopper, ...), a property that is necessary for the correct functioning of the KEM.

The library is integrated into OpenACC parallel regions via
\texttt{\#pragma acc routine}. Per-thread generator state is derived deterministically from a master seed, with one unique sub-seed per output element.

\subsection{Explicit Data Management}

Reliance on CUDA Unified Memory is avoided due to its non-deterministic
page-migration behavior under irregular access patterns. Device allocations are instead established once at program initialization via
\texttt{\#pragma acc enter data create(...)} directives.

All subsequent kernels operate on these device-resident buffers. The \texttt{present} clause is used to enforce consistency and suppress implicit data movement. Host-device synchronization is performed only at two well-defined points:
prior to CPU-side cryptographic hashing (device $\rightarrow$ host) and after
CPU-side message generation (host $\rightarrow$ device). This approach ensures
that transfers are initiated only when necessary and with strictly bounded scope.

\subsection{Kernel Parallelization Strategy}
\label{sec:kernels}

To maximize performance, all the operations described in Subsection~\ref{bottlenecks} are executed directly on the GPU. In the following, we briefly describe the main optimization strategies adopted in our implementation and the phases of the scheme that were primarily parallelized. Additional implementation details and kernel-level optimizations are available in the public repository accompanying this work.

We recall that, in KEM constructions, the encapsulation and decapsulation procedures rely on the encryption and decryption algorithms as subroutines for the underlying PKE scheme. Consequently, the optimization of the PKE phases (key generation, encryption, decryption) directly impact the KEM performance.
In the following, we describe how these phases were parallelized to improve the overall performance of the scheme.

\textbf{Key Generation:} Secret vector sampling from the centered binomial distribution $\chi^n_3$ is fully parallelized, since each component can be generated independently and directly mapped to GPU threads. The computation of the public vector $t = As + e$ assigns one row per thread. The corresponding inner products are evaluated sequentially within each thread in order to preserve deterministic generation of $A_{i,j}$ directly on the on the GPU from a seed. A fully vectorized implementation of this loop was intentionally avoided to prevent redundant recomputation overhead.

\textbf{Encryption and Decryption:} The encapsulation phase processes all 256 message bits within a single GPU kernel. Parallelism is exposed hierarchically by assigning independent message encryptions across OpenACC gangs; while the ciphertext components are computed concurrently across vector lanes.

For a lattice dimension of $n = 4096$, the total thread count per launch is more than a million concurrent GPU threads ($256 \times 4096 = 1\,048\,576$). 
The decapsulation phase follows an analogous strategy, where the computation of $\mu = v - s \cdot u \pmod{q}$ is parallelized using the same hierarchical decomposition and reduction-based accumulation scheme via an identical gang/vector decomposition with a parallel reduction.

\subsection{Performance Analysis}
\label{sec:perf_analysis}
System-level behavior was profiled using NVIDIA Nsight Systems 2025.2 with CUDA, NVTX, and OpenACC tracing enabled, while kernel-level microarchitectural characteristics were obtained using NVIDIA Nsight Compute with full metric coverage, including memory subsystem activity, warp scheduling behavior, and execution efficiency. This dual-level profiling enables a consistent separation between system-wide execution structure and low-level kernel behavior.

The workload, i.e. the number of keys generated, was fixed at $N = 10$ KEM instances with lattice dimension $n = 1024$, ensuring a controlled and representative comparison between implementations under identical input conditions.

\begin{figure*}[!t]
    \centering
    \subfloat[Nsight Systems timeline of the naive LWE-KEM implementation ($N=10$, $n=1024$) on the NVIDIA A100. Execution is dominated by host-side computation and PCIe data transfers, resulting in severe GPU starvation and a single-iteration time of approximately 2.787\,s.]{\includegraphics[width=\textwidth]{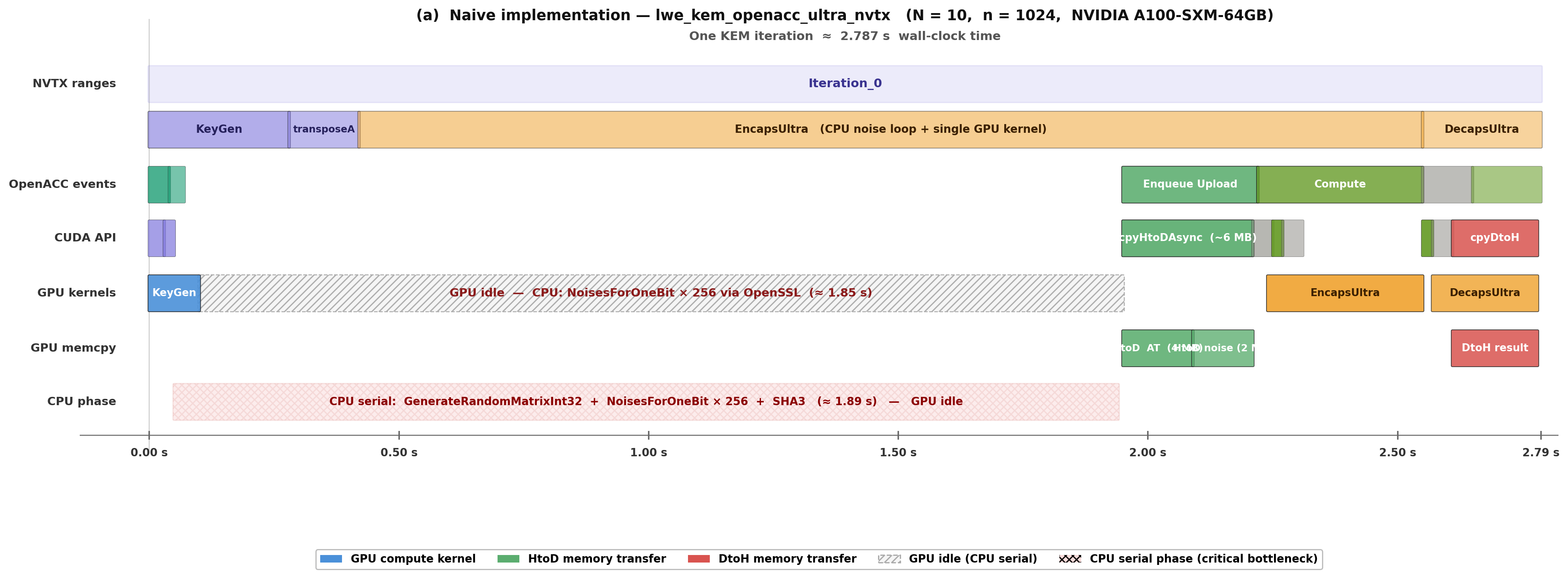}\label{fig:naive_timeline}}
    
    \vspace{2mm}
    
    \subfloat[Nsight Systems timeline of the optimized LWE-KEM implementation ($N=10$, $n=1024$) on the NVIDIA A100. Host-device communication is restricted to an 8-byte seed transfer, enabling sustained GPU activity ($\sim$98\% of iteration time) and reducing the single-iteration time to 8.24\,ms.]{\includegraphics[width=\textwidth]{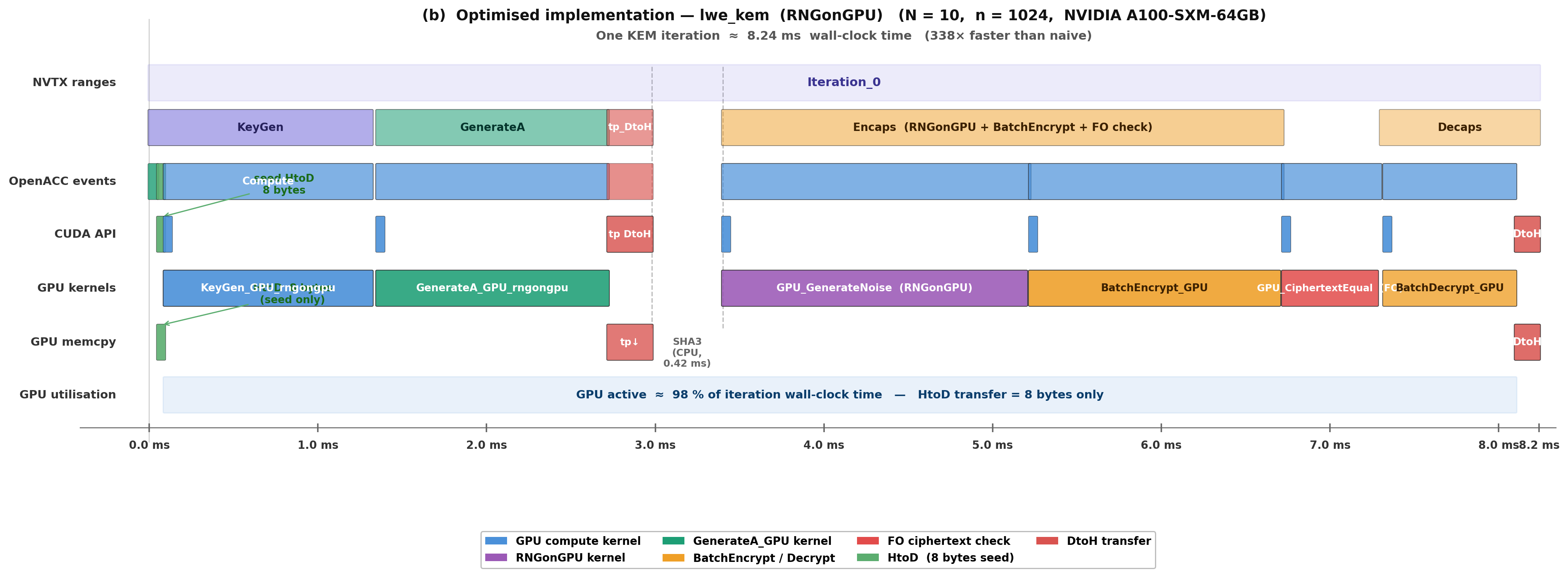}\label{fig:optimized_timeline}}
    
    \caption{Comparison of LWE-KEM execution timelines on the NVIDIA A100}
    \label{fig:timelines}
\end{figure*}

As shown in Figure \ref{fig:timelines}, a substantial performance improvement is observed in the optimized implementation. Total execution time is reduced from $27.869\,\mathrm{s}$ in the naive OpenACC baseline to $0.082\,\mathrm{s}$, corresponding to a speedup of approximately $338\times$. This improvement cannot be attributed to a single factor but emerges from the combined effects of reduced host-device communication, improved kernel-level parallelism, and more efficient utilization of GPU resources.

In the naive implementation, execution is largely dominated by host-side processing. Matrix generation and randomness construction are performed on the CPU prior to kernel invocation, during which the GPU remains idle for significant periods. In addition, for each encapsulation, large data structures, including public matrices and noise vectors, are transferred from host to device. These repeated transfers introduce substantial latency due to PCIe bandwidth constraints and lead to a fragmented execution timeline, where short kernel executions are interleaved with comparatively long CPU-bound phases.

In contrast, the optimized implementation shifts the majority of computation to the device. Only a small seed value is transferred from host to GPU, while matrix elements and random samples are generated directly on-device. As a result, repeated bulk transfers are eliminated, and kernel execution proceeds in a more continuous manner. Nsight Systems traces confirm sustained GPU activity with reduced idle intervals, indicating improved overlap of computation and memory operations at the system level.

Further insights are obtained from Nsight Compute metrics. The naive implementation exhibits low occupancy, with a limited number of active warps per streaming multiprocessor. This restricts the ability to hide memory latency effectively. Moreover, memory accesses are frequently uncoalesced, resulting in inefficient use of global memory bandwidth. Warp stall analysis indicates that the dominant source of inefficiency arises from memory dependency stalls, consistent with a memory-bound execution profile. The optimized implementation improves both occupancy and memory efficiency. A larger number of independent threads is exposed through batching and restructuring of computation, leading to higher warp activation and improved latency hiding. Memory access patterns are also significantly improved, with a higher proportion of coalesced transactions and better cache utilization. Consequently, effective memory throughput increases and global memory pressure is reduced.

A change in stall behavior is also observed. In the naive implementation, stalls are primarily attributed to memory latency. In the optimized version, the dominant stall reasons shift towards synchronization-related delays, suggesting that memory is no longer the primary bottleneck. Overall, the execution transitions from a memory-bound regime in the naive case to a more compute- and synchronization-limited regime in the optimized implementation.

\section{Results and discussion}
\label{sec:exp}
In this section, we present the results obtained by running the complete KEM workflow across OpenMP and OpenACC, focusing on the scalability between different GPUs.
The performance of the GPU-accelerated LWE-KEM implementation was evaluated against the multicore CPU version, corresponding to the \texttt{OpenACC} and \texttt{OpenMP} branches of our repository. The systems used to gather performance data for this section are detailed in Appendix~\ref{sec:appendix}. All performance experiments in this section, unless specified otherwise, were conducted on a containerized environment based on the NVIDIA HPC SDK image version 24.3. The Dockerfile used to build the container is available inside the \texttt{main} branch of the LWE-KEM Git repository.
All time-to-solution measurements (hereafter used synonymously with execution time) were performed using the C++ \verb|steady_clock| from the \verb|chrono| library with timing calls embedded in the application.

All experiments were conducted using the validated parameter sets discussed in Section~\ref{sec:implementation}, with identical parameters for the OpenMP and OpenACC implementations. Since both implementations perform equivalent modular arithmetic and noise sampling, any decryption failure is due to the chosen parameters rather than to hardware-specific effects.

\subsection{CPU and GPU scalability tests}
As established in Section~\ref{sec:implementation}, the dominant computational kernels scale quadratically with the LWE dimension $n$, making them the primary contributors to runtime for larger problem sizes. This behavior is consistent with the performance trends reported in Table~\ref{tab:performance}. As $n$ grows, the computational workload increases rapidly, allowing the GPU implementation to better leverage its massive parallelism and high memory bandwidth. Consequently, the OpenACC implementation exhibits superior scalability compared to the OpenMP implementation, achieving higher speedups for larger problem sizes. Both the OpenMP and OpenACC benchmarks were executed on the same GH200 compute node to ensure a fair comparison under identical hardware and software conditions. The OpenMP experiments were performed using 72 threads, corresponding to all physical cores available on the Grace CPU.

\begin{figure}[h!]
\centering
\includegraphics[scale=0.8]{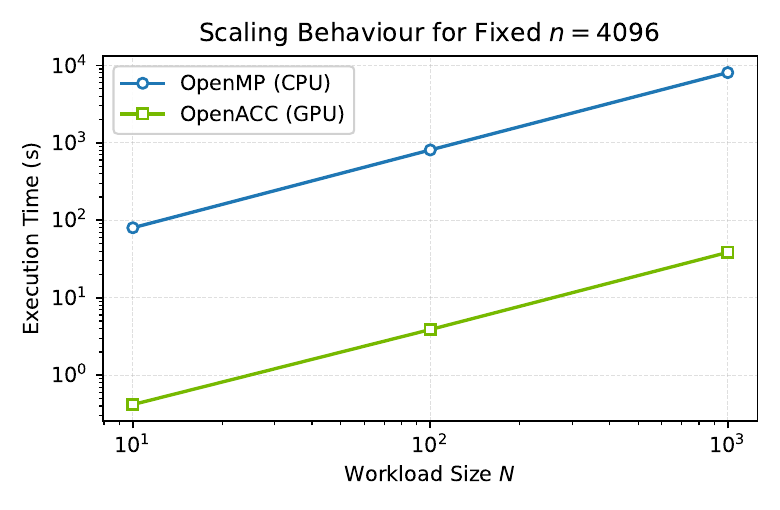}
\caption{Scaling behavior for fixed $n=4096$, showing execution time as a function of workload size N for OpenMP (CPU, blue) and OpenACC (GPU, green) implementations.} 
\label{fig1}
\end{figure}

\begin{table}[tp!]
\caption{Performance comparison between the multicore OpenMP implementation (72 threads) and the OpenACC implementation on the GH200 for $N = 100$ iterations. The table reports the average execution times of the KeyGen, Encaps, and Decaps phases (in $\mu s$), as well as the total workflow runtime (in $s$). The performance gain of OpenACC over OpenMP is quantified as $\mathrm{Speedup} = T_{\mathrm{OpenMP}} / T_{\mathrm{OpenACC}}$, which increases with $n$.}
\vspace{2mm}
\label{tab:performance}
\centering
\small

\setlength{\tabcolsep}{3pt}
\renewcommand{\arraystretch}{1.0}

\resizebox{\linewidth}{!}{%
\begin{tabular}{
c 
c c c c 
c c c c 
c}

\toprule
& \multicolumn{4}{c}{\textbf{OpenMP}} 
& \multicolumn{4}{c}{\textbf{OpenACC}} 
& \textbf{} \\
\cmidrule(lr){2-5} \cmidrule(lr){6-9}

\textbf{$n$} 
& \textbf{KeyGen} & \textbf{Encaps} & \textbf{Decaps} & \textbf{Total} 
& \textbf{KeyGen} & \textbf{Encaps} & \textbf{Decaps} & \textbf{Total} 
& \textbf{Speedup} \\

\midrule

32 & 881.83 & 274.04 & 147.49 & 0.13 & 3199.23 & 5740.13 & 5575.59 & 1.46 & 0.09 \\
128  & 7914.52	& 674.41 & 552.98 & 0.92 & 3212.22 & 5952.51 & 5579.62 & 1.48 & 0.62 \\
256  & 30232.35 & 1490.97 & 1375.53 & 3.32 & 3218.47 & 6258.3 & 5582.86 & 1.51 & 2.2 \\
512  & 116923.74 & 4064.75 & 3958.97 & 12.52 & 3305.69 & 7176.59 & 5997.84 & 1.65 & 7.57 \\
1024 & 477595.27 & 12902.04	& 12750.04 & 50.37 & 3636 & 8121.1 & 5785.05 & 1.76 & 28.61 \\
2048 & 2001914.9 & 46668.94 & 46803.2 & 209.63 & 4622.99 & 11175.6 & 6755.38 & 2.26 & 92.68 \\
4096 & 7742431.87 &	171860.34 & 172880.33 & 809.01 & 8576.53 & 19668.7 & 10551.8 & 3.89 & 208.13 \\

\bottomrule
\end{tabular}%
}
\end{table}
Figure~\ref{fig1} presents the scaling behavior for a fixed problem size of $n=4096$. The execution time increases approximately linearly with respect to the workload parameter $N$ for both implementations.
The OpenMP-based CPU implementation (blue) consistently exhibits higher execution times, whereas the OpenACC-based GPU implementation (green) maintains substantially lower values across all tested workloads. The roughly constant separation between the curves suggests that the relative speedup remains stable as $N$ increases. These results indicate that the GPU implementation consistently outperforms the CPU implementation, with both implementations exhibiting approximately linear scaling with respect to the workload size. 

\subsection{GPU performance}
\begin{figure}[t]
\centering
\includegraphics[scale=0.25]{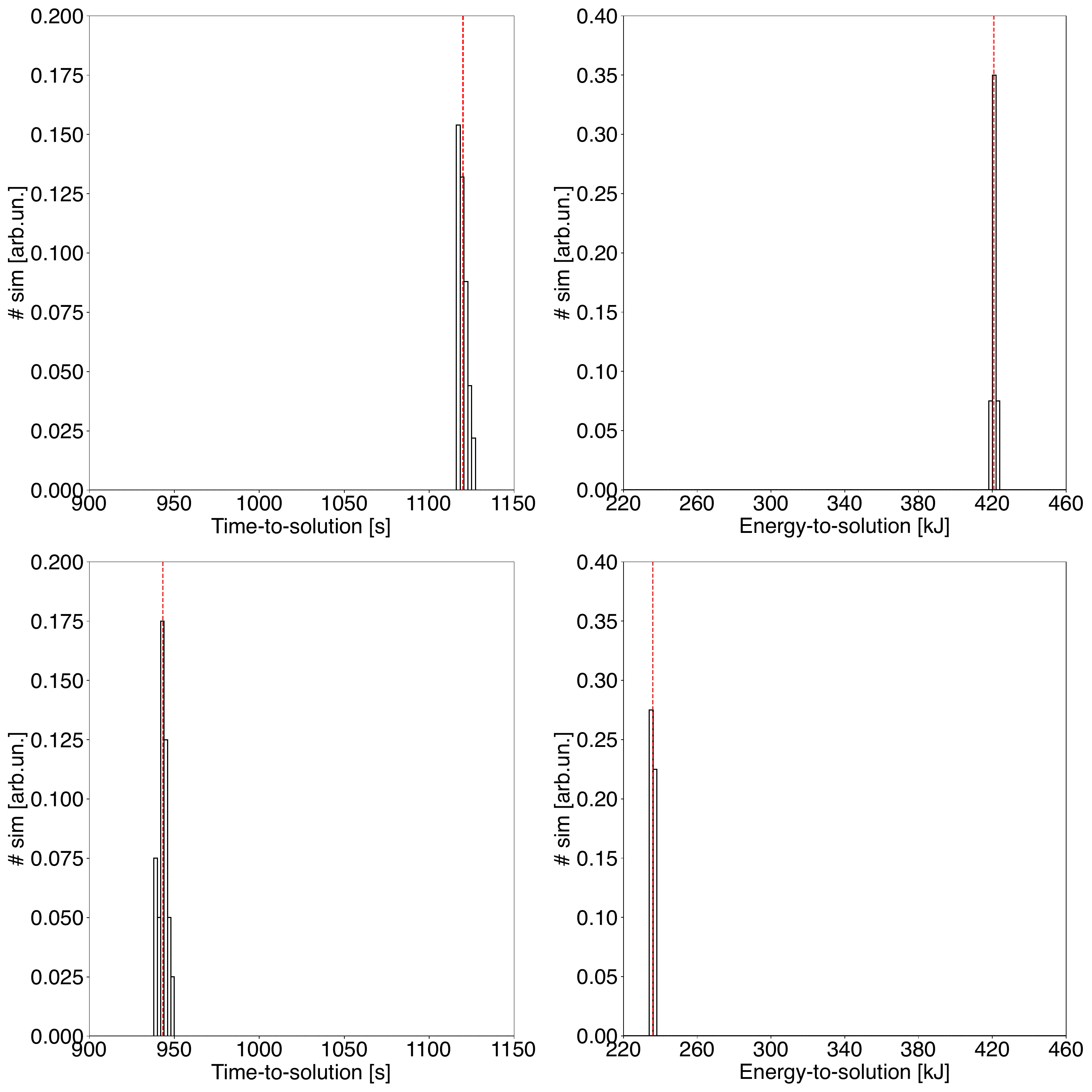}



\caption{
Top row: distribution of time-to-solution (TTS) over 20 simulations for runs with $n = 32768$ and $N = 1000$ on the x86 + Hopper system (left) and the GH200 node (right). 
Bottom row: distribution of energy-to-solution (ETS) over 20 simulations for the same runs and systems. 
The red dashed lines indicate the average TTS and ETS values for each system.
}
\label{fig:tts_ets}
\end{figure}

We assessed the performance of the optimized GPU implementation in terms of time-to-solution and energy-to-solution on x86 + Hopper and GH200 nodes featuring the most recent NVIDIA GPU models available to us.
For these tests we used $n=32768$ and $N=1000$, which correspond to the largest problem sizes that our GPUs could accommodate.
On both systems, the code was compiled using the NVHPC 25.7 compiler suite. CUDA 12.9 was used for the RNG library together with GCC 8.5.0 on the x86 + Hopper system and GCC 11.3.0 on GH200.
Each simulation was repeated 20 times to obtain sufficient statistics, with runs executed in batches and separated by 300-second pre- and post-run sleep intervals to ensure the system returned to idle between simulations. 
The top row of Figure~\ref{fig:tts_ets} shows the resulting time-to-solution distributions over 20 runs on the x86 + Hopper system (left) and the GH200 node (right). The reported times do not include the pre- and post-run sleep intervals.
On the x86 + Hopper system, the average execution time was $1119.9 \pm 2.6\,\mathrm{s}$, while on Grace Hopper it decreased to $943.2 \pm 2.4\,\mathrm{s}$, yielding a speedup of $1.19\times$. This improvement is likely due to the Grace Hopper Superchip unified memory design and higher memory bandwidth.
Energy-to-solution was computed by summing the energy consumed by the GPU and the CPU during each simulation, excluding the sleep intervals. For both systems, GPU energy consumption was estimated from the power measurements provided by \verb|nvidia-smi|. Following the methodology described in~\cite{amati,boella}, GPU power was sampled in the background from user space throughout each simulation and logged to a text file at a frequency of $1\,\mathrm{Hz}$. The GPU energy was then obtained by numerically integrating the power measurements over the simulation time.
For the x86 + Hopper system, CPU energy consumption was measured using \verb|perf stat -a -e| in combination with a one-second sleep interval, following the approach adopted in~\cite{Almerol1,Almerol2}. On GH200, CPU energy was estimated through system-level power monitoring interfaces. CPU power was obtained by reading the file \verb|/sys/class/hwmon/hwmon3/device/power1_average| once per second. 
This file reports the output of the sensor placed on the Grace Power Socket 0, averaged over a time window of $50\,\mathrm{ms}$. CPU energy was then computed by numerically integrating the measured power over time. As for the GPU measurements, polling of the CPU power data was performed in the background from user space during the simulations.
The distribution of energy-to-solution over 20 simulations for both systems is shown in the bottom row of Figure~\ref{fig:tts_ets}. On the x86 + Hopper system, a simulation consumed $420.9 \pm 1.0\,\mathrm{kJ}$, whereas the same simulation on GH200 required only $235.9 \pm 0.5\,\mathrm{kJ}$, corresponding to an energy saving factor of $1.78\times$. The reduction in energy consumption, which exceeds the observed time-to-solution improvement, is primarily attributable to the substantially lower CPU power consumption of the ARM-based Grace architecture.

\subsection{Cross-Generational GPU Architectural Scaling}
\label{sec:archit}
\begin{figure}[h!]
\centering
\includegraphics[scale=0.5]{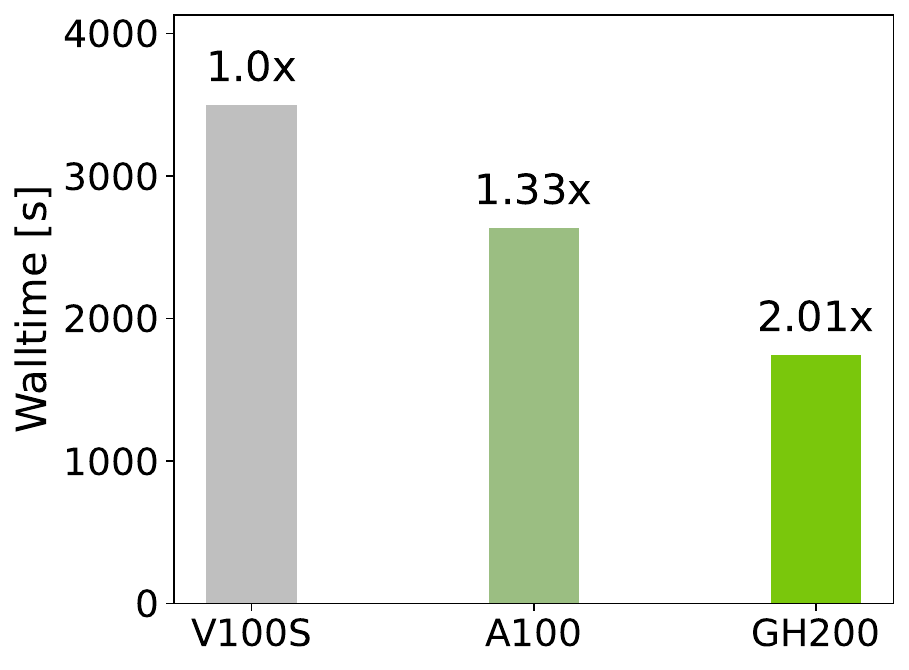}
\caption{Execution times and relative speedups for the different GPU models evaluated in our cross-generational comparison ($n=32768$, $N=1000$).} 
\label{gpu:scaling}
\end{figure}

Figure~\ref{gpu:scaling} presents the total time-to-solution of the LWE-KEM implementation for the largest configuration ($n = 32768$, $N = 1000$) across three NVIDIA GPU generations: Volta (V100S), Ampere (A100), and Hopper (GH200). The OpenMP-based CPU implementation was not able to complete this configuration within a 24-hour time limit on the GH200 node, due to cache contention, Non-Uniform Memory Access (NUMA) penalties, and OpenMP barrier overhead. 
All GPU experiments were executed in containers to ensure reproducible and consistent comparisons across heterogeneous platforms.

Execution times were $3500.08\,\mathrm{s}$ for V100S, $2634.69\,\mathrm{s}$ for A100, and $1740.39\,\mathrm{s}$ for GH200, yielding speedups of $1.33\times$ and $2.01\times$ for A100 and GH200, respectively, relative to the V100S baseline. To better understand these gains, the results were analyzed with respect to three main hardware characteristics: HBM memory bandwidth, cache hierarchy, and host--device communication architecture. Table~\ref{tab:gpu_architecture} summarizes the relevant architectural parameters of the evaluated platforms.

\begin{table}[htb!]
\caption{Key architectural characteristics of the GPUs evaluated in the cross-generational comparison.}
\label{tab:gpu_architecture}
\vspace{2mm}

\centering
\small

\setlength{\tabcolsep}{3pt}
\renewcommand{\arraystretch}{0.95}

\resizebox{0.7\linewidth}{!}{%
\begin{tabular}{
l
c c c
}

\toprule

\textbf{Parameter} 
& \textbf{V100S} 
& \textbf{A100} 
& \textbf{GH200} \\

\midrule

Microarchitecture        & Volta                & Ampere               & Hopper \\
HBM type                 & HBM2                 & HBM2e                & HBM3 \\
HBM bandwidth            & 1134\,GB/s           & $\sim$1555\,GB/s     & $\sim$3352\,GB/s \\
L2 cache                 & 6\,MB                & 40\,MB               & 60\,MB \\
SM count                 & 80                   & 108                  & 132 \\
CPU--GPU interconnect    & PCIe 3.0 $\times$16 & PCIe 4.0 $\times$16 & NVLink~4 \\
Interconnect bandwidth   & $\sim$16\,GB/s       & $\sim$32\,GB/s       & 900\,GB/s \\

\bottomrule

\end{tabular}%
}
\end{table}



Nsight Compute profiling (Subsection~\ref{sec:perf_analysis}) indicates that the LWE-KEM kernel is memory-bound, with stalls primarily driven by long-latency memory dependencies. Consequently, cross-generational performance is expected to scale predominantly with HBM bandwidth rather than computational throughput. The V100S provides 1134\,GB/s (HBM2), the A100 provides approximately 1555\,GB/s (HBM2e), and the GH200 provides approximately 3352\,GB/s (HBM3), giving bandwidth scaling factors of 1.37$\times$ and 2.96$\times$ relative to the V100S. The measured speedup on A100 (1.33$\times$) closely matches this scaling, indicating that memory throughput remains the primary constraint. The GH200 achieves 2.01$\times$, which is below the expected 2.96$\times$, suggesting that additional non-bandwidth-limited effects emerge at higher performance levels. In particular, on-demand matrix element generation introduces increasing arithmetic overhead as bandwidth improves, shifting part of the workload away from a purely memory-bound regime, while CPU-side costs from the Fujisaki--Okamoto transform impose a fixed serial component that limits end-to-end scaling under Amdahl’s law~\cite{Gustafson2011}.

For $n = 32768$, the working set is approximately 100\,MB, comprising large noise buffers and ciphertext data, which exceeds the L2 cache capacity across all architectures (6\,MB on V100S, 40\,MB on A100, and 60\,MB on GH200). As a result, capacity misses to HBM are unavoidable, although larger caches on A100 and GH200 reduce traffic for frequently reused structures such as the public vector $t$ and tiled regions of noise buffers, improving effective bandwidth use. The public matrix is generated on demand via a seed-expansion function rather than stored explicitly, avoiding multi-gigabyte allocations but introducing $\mathcal{O}(n^2)$ arithmetic overhead from per-element computation. This cost becomes increasingly significant as memory bandwidth improves, contributing to sublinear scaling on newer architectures. Additionally, host-device synchronization associated with cryptographic hashing persists across generations, with interconnect improvements (PCIe 3.0, PCIe 4.0, and NVLink 4) reducing but not eliminating transfer overhead. 

Overall, the cross-generational scaling of LWE KEM performance is governed by the interplay of three factors: (i) HBM bandwidth improvements, which drive the primary gains for the memory-bound GPU kernels; (ii) growing relative compute overhead from on-demand matrix generation, which attenuates bandwidth-driven scaling at higher performance levels; and (iii) fixed CPU-side costs from the Fujisaki-Okamoto transform, which establish a diminishing-returns ceiling via Amdahl's law. These results highlight that for memory-bound crypto workloads with significant non-GPU components, bandwidth improvements give meaningful but eventually saturating gains, and full utilization of next-generation architectures requires co-optimization of the full heterogeneous pipeline.

\section{Conclusion and Future work}
\label{sec:confun}
This work presents a KEM based on plain LWE and describes our strategy to accelerate the algorithm using OpenACC.
Unlike structured variants such as Ring-LWE and Module-LWE, plain LWE avoids additional algebraic structure beyond standard linear algebra over finite fields.
While this choice significantly increases computational requirements, it also relies on more conservative security assumptions.
The proposed implementation leverages the inherent parallelism of operations such as matrix--vector products, combining intra-operation batching, explicit data management, and on-device randomness generation to reduce host-device communication overhead and expose large-scale fine-grained parallelism on different GPU architectures.
In addition, by integrating the RNGonGPU library, we generate deterministic, cryptographically secure pseudorandom strings directly on the GPU, further limiting data transfers.
Indeed, profiling results (Figure~\ref{fig:naive_timeline}) show that our initial naive implementation without RNGonGPU and the rest of the optimizations is dominated by CPU-side noise generation and host-device coordination overheads.
In the enhanced implementation (Figure~\ref{fig:optimized_timeline}) we managed to reduce memory-transfer overheads, enabling more sustained GPU activity while shifting the performance bottleneck towards synchronization and coordination of host-side operations.

Node-to-node benchmarks comparing the OpenMP-parallelized CPU reference implementation with the accelerated GPU code show speedups of up to $208\times$ on the GH200 platform, depending on the value of $n$ (Table~\ref{tab:performance}). In particular, the GPU implementation begins to outperform the multicore CPU baseline for $n \geq 256$, where kernel parallelism and memory throughput are sufficient to amortize data-management overheads (Table~\ref{tab:performance}).
In addition, the proposed approach enabled the execution of large parameter configurations that are impractical on CPUs due to memory contention, synchronization overhead, and limited scalability.

We further evaluated the accelerated implementation in terms of execution time and energy consumption across different GPU models and architectures. In particular, the GH200 platform achieves up to a $1.78\times$ reduction in energy consumption compared to the x86 + Hopper configuration, whereas the reduction in execution time is more modest, reaching at most $1.19\times$.

A cross-generation analysis across Volta, Ampere, and Hopper GPUs confirms that memory bandwidth remains the primary limiting factor for plain LWE workloads, while CPU-side operations increasingly constrain scalability as parameters grow.


Overall, our results show that the proposed LWE-based KEM and its directive-based GPU implementation provide a practical and efficient path toward scalable post-quantum cryptography on modern heterogeneous platforms, possibly shifting the practicality boundary of conservative plain LWE cryptography.

Future work will investigate OpenMP portability across heterogeneous accelerator architectures, particularly AMD and Intel GPUs, to assess its effectiveness on different architectures, as C/C++ OpenACC support for AMD GPUs still lacks mature compilers. We will also extend the implementation to multi-GPU and distributed HPC environments for larger workloads and higher throughput. Additionally, further security and side‑channel analyses will be conducted to evaluate robustness in practical deployments.

\begin{credits}
\subsubsection{\discintname}


The authors have no competing interests to declare that are relevant to the content of this article.
T.L. developed the OpenMP-parallelized CPU implementation.
T.L. and N.S. contributed equally to the GPU porting and optimization of the application.
S.R. conducted the GPU experiments in containerized environments. 
E.B. carried out the energy measurements.
All authors contributed to the conception of the study, the analysis and interpretation of the results, and the writing and revision of the manuscript.

During the preparation of this article, artificial intelligence tools (ChatGPT and Grammarly) were used exclusively to assist with writing and language editing. All scientific content and analyses are the responsibility of the authors.
\end{credits}

\appendix
\section{Testbed Details}
\label{sec:appendix}
Our work was developed and tested on four compute nodes containing Volta-, Ampere-, and Hopper-generation NVIDIA GPUs.
The Volta-based platform consists of two AMD EPYC 7282 processors (SMT enabled, 2.8\,GHz), 512\,GB DDR4 memory (3200\,MT/s), and two NVIDIA V100S GPUs with 32\,GB HBM2 (High Bandwidth Memory) each connected through PCIe 3.0 $\times$16. The system runs Red Hat Enterprise Linux 9.4 with NVIDIA driver version 545.23.08 and utilizes Podman 5.2.2 for containerized execution. The Ampere-based platform, hosted on the Booster partition of the \textit{Leonardo} supercomputer~\cite{Leonardo}, features an Intel Xeon Platinum 8358 processor (SMT disabled, 2.6\,GHz), 512\,GB DDR4 memory, and four NVIDIA A100 GPUs with 64\,GB HBM2e each connected through PCIe 4.0 $\times$16. The platform operates on Red Hat Enterprise Linux 8.8 and executes workloads through SingularityPRO 4.3.1. The x86 Hopper platform integrates dual AMD EPYC 9554 processors (SMT disabled, 3.10\,GHz), 512\,GB DDR5 memory (4800\,MT/s), and one NVIDIA H100 PCIe GPU with 80\,GB HBM3 connected through PCIe 5.0 $\times$16. It runs Red Hat Enterprise Linux 8.6. Finally, the Grace Hopper platform employs the NVIDIA GH200 Grace Hopper Superchip, combining an ARM Neoverse V2 CPU (SMT disabled, 3.47\,GHz), 480\,GB LPDDR5 memory, and a Hopper GPU with 96\,GB HBM3 memory. The CPU and GPU are connected through NVLink~4, providing up to 900\,GB/s bidirectional bandwidth. This system runs Ubuntu 22.04.5 with Docker 29.0.2 for containerized execution. Across all evaluated systems, the Linux \textit{performance} CPU frequency governor was enforced to ensure stable measurements.

\bibliographystyle{splncs04}
\bibliography{./bib/refs}
\end{document}